\documentclass[pdflatex,sn-mathphys-ay]{sn-jnl}

\usepackage{graphicx}
\usepackage{amsmath,amssymb,amsfonts}
\usepackage[title]{appendix}
\usepackage{booktabs}
\usepackage{array}
\usepackage{tabularx}
\usepackage{makecell}
\usepackage{subcaption}
\usepackage{url}
\usepackage{xcolor}
\usepackage{ragged2e}
\usepackage{placeins}
\usepackage{float}

\newcolumntype{Y}{>{\RaggedRight\arraybackslash}X}
\newcolumntype{C}{>{\centering\arraybackslash}X}

\newcommand{\mphi}{m_{\phi}}
\newcommand{\logmphi}{\log_{10}(\mphi/\mathrm{eV})}
\newcommand{\mtt}{m_{22}}
\newcommand{\Mvir}{M_{\rm vir}}

\newcommand{\Mstar}{M_{\star}}
\newcommand{\rc}{r_c}
\newcommand{\rhoc}{\rho_c}
\newcommand{\Msun}{M_{\odot}}
\newcommand{\Upsdisk}{\Upsilon_{\rm disk}}
\newcommand{\Upsbul}{\Upsilon_{\rm bul}}
\newcommand{\rhat}{\hat{r}}
\newcommand{\kms}{\,\mathrm{km\,s^{-1}}}
\newcommand{\kpc}{\,\mathrm{kpc}}
\newcommand{\rhocrit}{\rho_{\rm crit}}

\providecommand{\doi}[1]{doi: \url{https://doi.org/#1}}

\begin{document}

\title[Joint ULDM population inference]{A reusable hierarchical framework for joint inference of ultralight-dark-matter mass and core-halo scaling}

\author*[1]{\fnm{Prasun} \sur{Panthi}}\email{ppanthi27@wabash.edu}
\author[1]{\fnm{Md Shahrier Islam} \sur{Arham}}
\affil*[1]{\orgdiv{Department of Physics}, \orgname{Wabash College},
\city{Crawfordsville}, \state{Indiana}, \country{USA}}

\abstract{Ultralight-dark-matter rotation-curve analyses often infer the particle mass after fixing a relation between the central soliton and its host halo. The resulting mass constraint is then conditional on a population-level relation that the data may not independently support. We present a reusable hierarchical Bayesian framework that instead infers the ultralight particle mass and the core-halo scaling exponent jointly from a galaxy population. The differentiable forward model combines a Schive-normalized soliton with a smoothly matched, regularized NFW envelope and allows galaxy-level halo and stellar parameters to be inferred together with the global dark-matter parameters. We apply the framework to 106 SPARC galaxies, including 26 systems with bulges, and sample the resulting 346-dimensional posterior with JAX/NumPyro NUTS. The free-scaling run has zero divergences and \(\rhat\simeq1.000\) for the global parameters. The posterior moves to the high-mass, weak-scaling boundary, with \(\logmphi=-19.20^{+0.12}_{-0.11}\) and \(\alpha=0.014^{+0.023}_{-0.011}\). In this regime, the solitonic cores lie below the radial scales probed by the rotation curves, while the baryonic terms and outer NFW envelope carry the visible fits. The same boundary behaviour remains after removing UGC06787 and after widening the high-mass stellar-to-halo-mass prior. The selected SPARC sample therefore does not give an interior joint constraint on the particle mass and core-halo scaling relation within the adopted model. The framework makes this unresolved-core limit explicit instead of interpreting the prior boundary as an interior mass constraint. Its modular structure also allows the same analysis to be applied to synthetic recovery tests, expanded or higher-resolution rotation-curve samples, and alternative halo models.}


\maketitle

\section{Introduction}
\label{sec:intro}

Galaxy rotation curves are a direct probe of the mass distribution in disk galaxies. The observed circular velocity at each radius depends on the total enclosed mass, including stars, gas, and any dark component. Rotation curves therefore provide a useful test of how baryonic matter and dark matter combine to shape galactic gravitational potentials.

The SPARC database provides a standard sample for this type of test. It contains rotation curves for nearby disk galaxies together with baryonic velocity components from gas, stellar disks, and bulges where present \citep{Lelli2016}. These decomposed contributions make SPARC useful for fitting dark-matter models because the baryonic and non-baryonic parts of the rotation curve can be treated separately within the same forward model.

Ultralight or fuzzy dark matter models predict structure on galactic scales that differs from standard cold dark matter. In these models, the wave nature of the dark matter can produce a central solitonic core surrounded by an outer halo \citep{Hu2000,Schive2014Nature,Hui2017}. Rotation curves can therefore test whether a soliton-like central component is supported by galaxy data, and how that component connects to the outer halo. For broad reviews of dark matter and ultralight/fuzzy dark matter, including observational constraints, halo density profiles, and galaxy-scale tests, see \citet{Cirelli2026Review} and \citet{EberhardtFerreira2025Review}.

A key modelling choice is the relation between the soliton core and the host halo. Many rotation-curve analyses fix this relation before fitting the data. This reduces the number of free parameters, but the inferred particle mass is then conditional on the chosen relation. It does not show whether the galaxy population supports that relation independently. In this work we take a different approach. We treat the core-halo scaling exponent as a global parameter and infer it jointly with the ultralight particle mass. In the notation used below, these parameters are \(\alpha\) and \(\mphi\). The data can then determine whether both parameters have an interior population-level solution or whether the model moves toward a limiting regime.

The direct comparator for our mass prior is the Bayesian SPARC analysis of \citet{Khelashvili2023}. They use \(\log_{10}(m_{22})\in[-3,3]\), equivalent to \(\logmphi\in[-25,-19]\). We adopt this range for the runs. The lower end includes the ultralight masses commonly discussed in the FDM literature. The upper end reaches the high-mass regime where the soliton becomes compact on observed rotation-curve scales and the model approaches a CDM-like limit. Lyman-alpha forest studies give strong lower limits on FDM masses under cosmological assumptions \citep{Irsic2017,Rogers2021}. We cite those bounds for context. We do not use them as priors in the SPARC fit.

We build a hierarchical Bayesian pipeline for fitting a Schive-normalized soliton plus regularized NFW envelope to galaxy rotation curves. The model uses one global particle mass and one global core-halo scaling exponent. Each galaxy has its own halo mass, concentration, and stellar mass-to-light ratios. The transition between the soliton and the NFW envelope is differentiable, which allows the full population model to be sampled with Hamiltonian Monte Carlo. The global parameters, galaxy-level parameters, sample selection, and halo prescription are kept separate in the implementation. The same pipeline can therefore be reused with other galaxy samples, synthetic populations, and alternative halo models.

We apply the pipeline to 106 SPARC galaxies selected by inclination, quality flag, number of velocity points, and mean velocity uncertainty. The posterior has 346 sampled dimensions in the free-scaling run. We sample it with JAX/NumPyro NUTS and report convergence diagnostics, boundary diagnostics, stability checks, fixed-scaling sensitivity runs, and a representative rotation-curve decomposition.

The main result is that the free-scaling hierarchy converges to a boundary solution. Within the Schive-normalized model and the standard prior range \(\logmphi\in[-25,-19]\), the posterior reaches the high-mass edge of the particle-mass prior and the low-scaling edge of the core-halo exponent prior. In this regime the soliton core is compact on the observed rotation-curve scales, and the visible fit is carried mainly by the baryonic terms and the regularized NFW envelope. The same boundary behaviour is found after removing UGC06787 and after widening the high-mass stellar-to-halo-mass prior.

The purpose of this paper is to introduce the joint-inference framework and demonstrate it on SPARC. The boundary solution is not treated as a measurement at the edge of the particle-mass prior. It shows that, for the selected sample and adopted model, the data do not jointly identify an interior particle mass and core-halo scaling relation. When the proposed soliton structure is unresolved, the hierarchy makes the limiting regime explicit instead of reporting the boundary value as an interior constraint.

\noindent\textbf{Principal contributions.} The main contributions of this work are:
\begin{enumerate}
\item a differentiable hierarchical pipeline that infers the ultralight particle mass and core-halo scaling exponent jointly instead of fixing the scaling relation before fitting;
\item a separation of the two global dark-matter parameters from the galaxy-level halo and stellar parameters in a 346-dimensional population model;
\item an application to 106 SPARC galaxies that identifies a converged unresolved-core boundary solution when the population scaling is allowed to vary;
\item diagnostics that separate an interior population constraint from a prior-boundary solution and from modes produced conditionally at fixed scaling;
\item stability checks showing that the free-scaling boundary behaviour is unchanged after removing UGC06787 and after widening the high-mass stellar-to-halo-mass prior; and
\item a modular implementation that can be used for synthetic recovery tests, new rotation-curve samples, and alternative inner- and outer-halo models.
\end{enumerate}

The paper is organized as follows. Section~\ref{sec:theory} defines the soliton-plus-NFW forward model. Section~\ref{sec:data} describes the SPARC data and sample selection. Section~\ref{sec:model} gives the hierarchical Bayesian model, priors, likelihood, and sampling procedure. Section~\ref{sec:results} presents the posterior results and diagnostic checks. Section~\ref{sec:discussion} discusses the interpretation and limitations of the boundary solution. Section~\ref{sec:conclusion} summarizes the conclusions.

\section{Theoretical framework}
\label{sec:theory}

\subsection{Soliton inner profile}

The solitonic ground state of the Schr\"odinger-Poisson system has the empirical profile \citep{Schive2014PRL}
\begin{equation}
\rho_{\rm sol}(r)=\frac{\rhoc}{\left[1+0.091(r/\rc)^2\right]^8},
\label{eq:soliton_shape}
\end{equation}
where \(\rc\) is the core radius and \(\rhoc\) is the central density. The numerical constants \(0.091\) and \(8\) are empirical fits from cosmological FDM simulations \citep{Schive2014PRL} and are taken to be fixed in the present analysis. The Schive-normalized relation between central density, boson mass, and core radius is
\begin{equation}
\rhoc = \rho_{\rm norm}\,\mtt^{-2}\left(\frac{\rc}{1\kpc}\right)^{-4},
\label{eq:rhoc_schive}
\end{equation}
with \(\mtt=\mphi/(10^{-22}\mathrm{eV})\) and \(\rho_{\rm norm}=1.9\times10^7\,\Msun\,\mathrm{kpc}^{-3}\) \citep{Schive2014PRL,Khelashvili2023}. At fixed \(\rc\), increasing \(\mtt\) lowers the central density as \(\mtt^{-2}\). Combined with the core-radius relation below, increasing \(\mphi\) also makes the core smaller. Eliminating \(\rc\) between Eqs.~\ref{eq:rhoc_schive} and \ref{eq:rc_schive} gives \(\rhoc\propto \mtt^2 M_9^{4\alpha}\). The negative mass exponent in Eq.~\ref{eq:rhoc_schive} is the density scaling at fixed core radius; the positive combined exponent follows after the core radius is allowed to co-vary with \(\mphi\).

\subsection{Core-radius relation}

The core radius is parameterized as
\begin{equation}
\rc=A_{\rm ch}\,\mtt^{-1}M_9^{-\alpha},
\label{eq:rc_schive}
\end{equation}
where \(M_9=\Mvir/(10^9\Msun)\). We use \(A_{\rm ch}=1.6\kpc\), the Schive core-radius normalization for \(\mtt=1\) and \(M_9=1\) \citep{Schive2014PRL}. The exponent \(\alpha\) controls how core size changes with halo mass. The Schive simulations give \(\alpha=1/3\) for their core-halo relation \citep{Schive2014PRL}. Other simulations and analyses report a range of behaviour \citep{Mocz2017,Schwabe2016,Nori2021,Chan2022}. We sample \(\alpha\) as a global parameter of the hierarchy.

\subsection{Outer NFW envelope and smooth transition}

The outer halo follows an NFW form \citep{Navarro1996,Navarro1997},
\begin{equation}
\rho_{\rm NFW}(r)=\frac{\rho_s}{(r/r_s)(1+r/r_s)^2}.
\label{eq:nfw}
\end{equation}
The sampled parameters \(\Mvir\) and \(c_{200}\) set the NFW scale radius and scale density. We define
\begin{equation}
R_{200}=\left(\frac{3\Mvir}{4\pi\,200\rhocrit}\right)^{1/3},
\qquad
r_s=\frac{R_{200}}{c_{200}},
\end{equation}
and
\begin{equation}
\rho_s=\frac{\Mvir}{4\pi r_s^3\left[\ln(1+c_{200})-c_{200}/(1+c_{200})\right]}.
\end{equation}
We regularize the inner cusp before blending the envelope with the soliton,
\begin{equation}
\rho_{\rm env}(r)=\left[1-\exp\left(-\left(\frac{r}{r_\epsilon}\right)^2\right)\right]\rho_{\rm NFW}(r).
\label{eq:regularized_nfw}
\end{equation}
The regularizer suppresses the NFW cusp near the soliton region. The logistic blend below also suppresses the envelope at small radii, so the envelope is deliberately down-weighted twice near the core. This choice keeps the composite profile smooth and prevents the NFW cusp from dominating the central soliton region.

The transition radius \(r_\epsilon\) is selected for each galaxy by a differentiable soft-argmin search over a fixed logarithmic radius grid. The searched quantity is \(|\log\rho_{\rm sol}(r)-\log\rho_{\rm env}(r)|\). The soft-argmin temperature is \(\beta=12\). The composite density is
\begin{equation}
\rho_{\rm DM}(r)=B(r;r_\epsilon)\rho_{\rm sol}(r)+[1-B(r;r_\epsilon)]\rho_{\rm env}(r),
\label{eq:composite_density}
\end{equation}
with
\begin{equation}
B(r;r_\epsilon)=\frac{1}{1+\exp[k(r-r_\epsilon)]},
\end{equation}
where \(k=3\). This construction gives a smooth forward model for Hamiltonian Monte Carlo.

\subsection{Rotation velocity}

The enclosed dark mass is
\begin{equation}
M_{\rm DM}(<r)=4\pi\int_0^r \rho_{\rm DM}(r')r'^2\,dr'.
\label{eq:enclosed_mass}
\end{equation}
The numerical integral uses a hybrid radial grid with 256 logarithmically spaced points and 320 linearly spaced points. The grid resolves the compact core while retaining stable behaviour at large radii. The dark contribution to the rotation curve is
\begin{equation}
V_{\rm DM}^2(r)=\frac{G M_{\rm DM}(<r)}{r}.
\label{eq:dark_velocity}
\end{equation}
The total predicted velocity is
\begin{equation}
V_{\rm model}^2(r)=V_{\rm gas}|V_{\rm gas}|+\Upsdisk V_{\rm disk}^2(r)+\Upsbul V_{\rm bulge}^2(r)+V_{\rm DM}^2(r).
\label{eq:v_model}
\end{equation}
The term \(V_{\rm gas}|V_{\rm gas}|\) preserves the SPARC sign convention for gas contributions, where negative gas values indicate that gas lowers the circular speed contribution in the mass model. The stellar mass-to-light ratios \(\Upsdisk\) and \(\Upsbul\) are inferred for each galaxy.

\section{Data and selection}
\label{sec:data}

We use the SPARC rotation-curve database \citep{Lelli2016}. Each rotation-curve file gives radius, observed velocity, velocity uncertainty, gas velocity, disk velocity, and bulge velocity when present. The table metadata give distance, inclination, quality flag, and flat rotation velocity.

We apply four cuts. We require inclination \(i\geq30^\circ\), quality flag \(Q\neq3\), at least ten velocity points, and mean velocity uncertainty below \(10\kms\). The final sample contains 106 galaxies. Of these, 26 have nonzero bulge components. The selected flat velocities span \(33.7\kms\) to \(332.0\kms\).

For the stellar-to-halo-mass prior, we estimate disk luminosity by integrating the SPARC disk surface-brightness profile,
\begin{equation}
L_{\rm disk}=2\pi\int_0^{r_{\rm max}} r\,\Sigma_{\rm disk}(r)\,dr.
\end{equation}
The integral is evaluated with the trapezoidal rule on the supplied radial grid. Galaxies without a usable disk surface-brightness profile receive a luminosity floor of \(10^5L_\odot\). This makes the stellar-to-halo-mass prior effectively uninformative for those systems.

Galaxies have different numbers of rotation-curve points. The implementation pads all arrays to a common length and uses a Boolean mask in the likelihood. Padded entries never contribute to the log likelihood.

\section{Hierarchical Bayesian model}
\label{sec:model}

The hierarchy separates sample-level and galaxy-level parameters. The particle mass scale and the core-halo scaling exponent are shared by all galaxies. The halo mass, concentration, and stellar mass-to-light ratios are inferred separately for each galaxy. The likelihood compares the model velocity to the observed SPARC velocity at each valid radius, while the priors set the halo-mass scale, concentration relation, and stellar-to-halo-mass information.

The two global parameters are \(\logmphi\) and \(\alpha\). The runs use
\begin{equation}
\logmphi\sim U(-25,-19),\qquad \alpha\sim U(0,1.5).
\label{eq:global_priors}
\end{equation}
The \(\logmphi\) prior follows the SPARC FDM range used by \citet{Khelashvili2023}. The same range is used here for one global \(\mphi\) because the global parameter represents the sample-level boson mass in the same physical interval used for per-galaxy FDM mass inference. The \(\alpha\) prior covers the values used in simulation-based core-halo relations and allows the data to move toward weak halo-mass dependence.

Each galaxy has local parameters \(\log_{10}\Mvir\), \(c_{200}\), \(\Upsdisk\), and, for bulge galaxies, \(\Upsbul\). The free-\(\alpha\) dimension count is
\begin{equation}
2 + 80\times3 + 26\times4 = 346,
\end{equation}
where 80 galaxies have no bulge and 26 have a bulge. The fixed-\(\alpha\) runs have 345 sampled dimensions.

We use a \(V_{\rm flat}\to\Mvir\) prior derived from the NFW maximum-circular-velocity relation,
\begin{equation}
V_{\rm max}^2=\frac{G\Mvir}{R_{200}}\frac{0.216c_{200}}{\ln(1+c_{200})-c_{200}/(1+c_{200})}.
\end{equation}
We evaluate this relation at \(c_{200}=10\) and assign a scatter of 0.35 dex in \(\log_{10}\Mvir\). This is an approximate halo-mass prior tied to the observed flat velocity. The model still samples \(c_{200}\) separately, so the prior does not enforce exact self-consistency between \(V_{\rm flat}\) and the sampled concentration. We list this approximation as a modelling limit in Section~\ref{sec:discussion}. The concentration prior follows \citet{Dutton2014},
\begin{equation}
\log_{10} c_{200}\sim \mathcal{N}\left(0.905-0.101[\log_{10}(\Mvir/\Msun)-12],\,0.11\right).
\end{equation}
The scatter is applied in dex to \(\log_{10}c_{200}\), matching the implementation used in the runs.

We also apply a Moster stellar-to-halo-mass prior \citep{Moster2013},
\begin{equation}
\frac{\Mstar}{\Mvir}=\frac{2N}{(\Mvir/M_1)^{-\beta}+ (\Mvir/M_1)^{\gamma}}.
\end{equation}
The parameters are \(M_1=10^{11.59}\Msun\), \(N=0.0351\), \(\beta=1.376\), and \(\gamma=0.608\). We invert this monotonic relation with bracketed bisection in \(\log_{10}\Mvir\). The scatter is 0.20 dex for \(\log_{10}\Mstar\geq9.5\), 0.30 dex for \(9.0\leq\log_{10}\Mstar<9.5\), and disabled for \(\log_{10}\Mstar<9.0\).

The \(V_{\rm flat}\) prior and the SMHM prior are multiplied in the posterior when both are active. In log space, their log-density contributions add to the likelihood and other priors. When the SMHM prior is disabled, the \(V_{\rm flat}\) prior supplies the halo-mass scale by itself.

The likelihood is a masked Gaussian over all valid rotation-curve points,
\begin{equation}
\ln\mathcal{L}=-\frac{1}{2}\sum_i\sum_j \left[\frac{V_{\rm obs}^{(i,j)}-V_{\rm model}^{(i,j)}}{\sigma_V^{(i,j)}}\right]^2 + {\rm const.}
\end{equation}
The index \(i\) labels galaxies and \(j\) labels radii within each galaxy. The constant absorbs the Gaussian normalization term, which does not affect posterior geometry because the observational uncertainties \(\sigma_V^{(i,j)}\) are fixed by the data.

We sample with the No-U-Turn Sampler \citep{Hoffman2014} as implemented in NumPyro \citep{Phan2019}. The forward model is JIT-compiled with JAX \citep{Bradbury2018}. The runs use two chains, 1000 warmup steps, 2000 post-warmup samples per chain, target acceptance 0.95, and maximum tree depth 11. Both chains are initialized independently using NumPyro's \texttt{init\_to\_median} strategy from the prior. The two chains run in parallel across two NVIDIA RTX 4090 GPUs. Wall times are therefore hardware-specific.

We assess convergence with rank-normalized split \(\rhat\), bulk effective sample size, tail effective sample size, and NUTS divergence count \citep{Vehtari2021}. We report a posterior only when the global parameters have \(\rhat<1.01\), \(\mathrm{ESS}_{\rm bulk}>400\) per global parameter, and zero divergences. Runs that fail these criteria are reported as sampler diagnostics rather than posterior measurements.

\subsection{Reusable pipeline structure}

The inference is organized around a common population model. The global block contains \(\mphi\) and \(\alpha\). The local block contains the halo and stellar parameters for each galaxy. The data block supplies the observed rotation curve and baryonic velocity components. This separation allows the number of galaxies to change without changing the global model. It also allows mock galaxies to be passed through the same likelihood and diagnostics as the observed sample.

The density model is evaluated inside the same differentiable forward calculation. A different core profile, envelope, or transition rule can therefore be tested while retaining the hierarchical parameter structure and sampling workflow. This makes the pipeline useful both for new data and for injection-recovery studies in which the input particle mass and core-halo relation are known.

\section{Results}
\label{sec:results}

\suppressfloats[t]

\subsection{Free-\texorpdfstring{\(\alpha\)}{alpha} posterior}

The free-\(\alpha\) run converges cleanly. The run has zero divergences. The global parameters
have rank-normalized split \(\rhat\simeq1.000\). The bulk effective sample sizes are 1844 for
\(\logmphi\) and 3727 for \(\alpha\). The tail effective sample sizes are reported in
Table~\ref{tab:fixed_alpha}. Because the posterior sits on hard prior boundaries, quantiles
and boundary fractions are the primary summary.

\begin{figure}[!ht]
\centering
\includegraphics[width=0.72\linewidth]{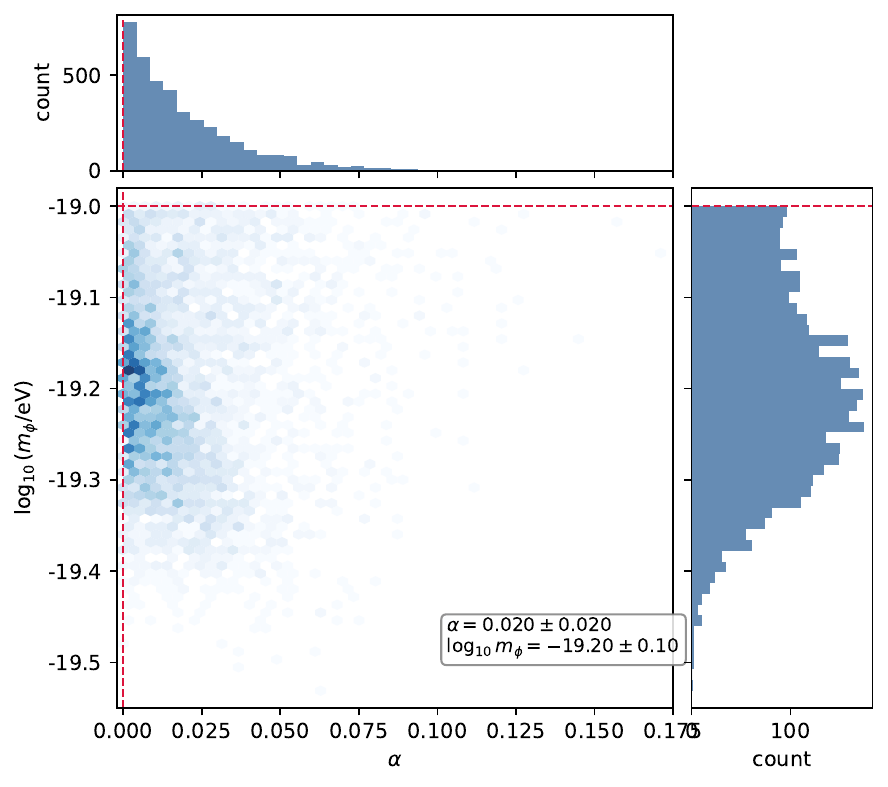}
\caption{Headline free-\(\alpha\) posterior. The plot shows the joint posterior in
\((\alpha,\logmphi)\) and the one-dimensional marginals from 4000 post-warmup samples.
Dashed red lines mark the standard prior boundaries. The posterior mass is concentrated
near the high-\(\mphi\) prior boundary and near \(\alpha=0\).}
\label{fig:posterior}
\end{figure}

\begin{figure}[!htbp]
\centering
\includegraphics[width=0.88\linewidth]{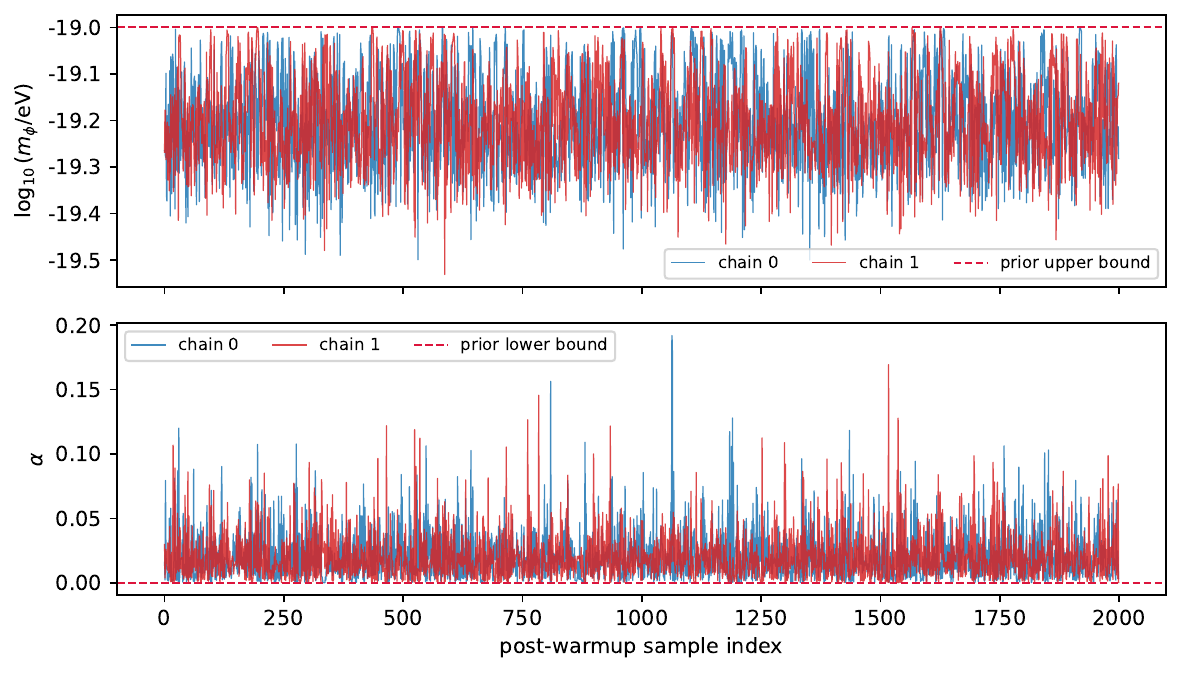}
\caption{Trace plots for the headline free-\(\alpha\) run. Both chains explore the same boundary region. The run has zero divergences and \(\rhat\simeq1.000\) for the global parameters.}
\label{fig:traces}
\end{figure}

The posterior medians and 16/84 percentile intervals are
\begin{equation}
\logmphi=-19.20^{+0.12}_{-0.11},
\end{equation}
\begin{equation}
\alpha=0.014^{+0.023}_{-0.011}.
\end{equation}
The corresponding 5/95 percentile intervals are \([-19.37,-19.03]\) for \(\logmphi\) and \([0.001,0.060]\) for \(\alpha\). Compact mean and standard-deviation descriptors are \(\logmphi=-19.20\pm0.10\) and \(\alpha=0.020\pm0.020\). Figure~\ref{fig:posterior} shows the joint posterior. Figure~\ref{fig:traces} shows the chain traces.

The ESS criteria reported here apply to the global parameters. Per-galaxy parameter ESS values were not individually audited; a systematic per-galaxy convergence check is deferred.

The posterior is boundary-pinned. It reaches the upper edge of the standard \(\logmphi\) prior and the lower edge of the \(\alpha\) prior. The quantile diagnostics make this explicit: 9.5\% of \(\logmphi\) samples lie within 0.05 dex of the upper prior bound, 20.5\% lie within 0.10 dex, 22.2\% of \(\alpha\) samples lie below 0.005, and 38.5\% lie below 0.010. The posterior occupies the region where the soliton core is compact and the core-radius dependence on halo mass is weak.

This is the headline result. Under the Schive normalization and the standard literature prior, the hierarchical SPARC fit identifies a boundary solution. Within the adopted Schive-normalized forward model and the standard prior range, the free-\(\alpha\) hierarchy does not identify an interior soliton component.

\subsection{Galaxy-exclusion check}

We reran the free-\(\alpha\) pipeline after removing UGC06787. UGC06787 is a bulge galaxy in this selection, so the run used 105 galaxies, including 25 with bulges, and sampled 342 dimensions. It converged cleanly with zero divergences. The posterior summaries are
\begin{equation}
\logmphi=-19.20\pm0.11,
\qquad
\alpha=0.020\pm0.019.
\end{equation}
The posterior is not changed by removing UGC06787. The boundary behaviour survives the galaxy-exclusion run.

\subsection{SMHM-prior widening check}

\begin{figure}[!htbp]
\centering
\includegraphics[width=0.82\linewidth]{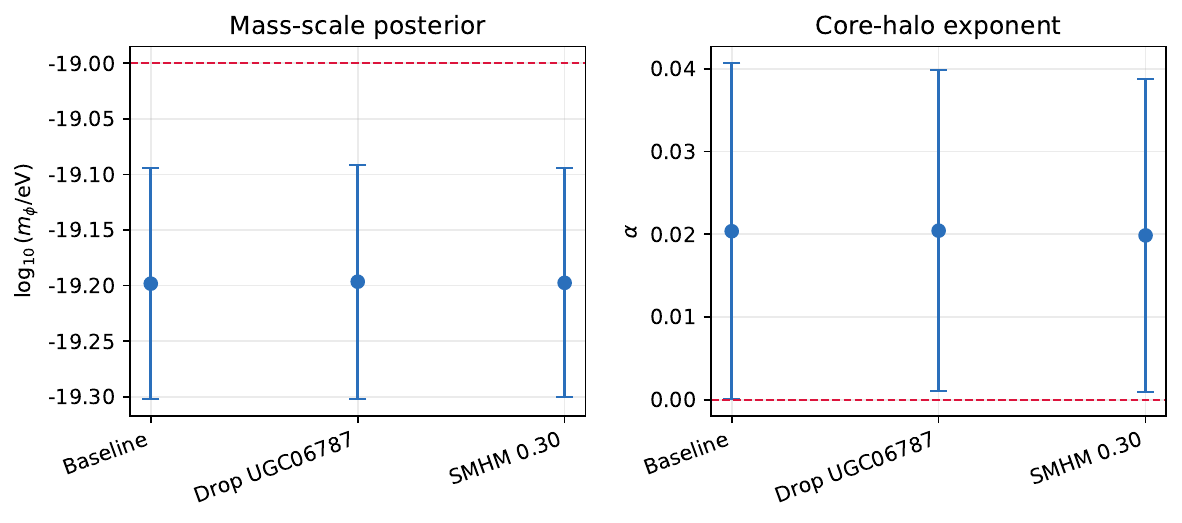}
\caption{Stability checks. Error bars show one posterior standard deviation for the baseline free-\(\alpha\) run, the run excluding UGC06787, and the run with high-mass SMHM scatter widened from 0.20 dex to 0.30 dex. The three free-scaling configurations return the same boundary solution at the quoted precision. The figure is a compact summary; the quantitative diagnostics are listed in Table~\ref{tab:fixed_alpha}.}
\label{fig:robustness}
\end{figure}

We also reran the free-\(\alpha\) pipeline after increasing the high-mass stellar-to-halo-mass scatter from 0.20 dex to 0.30 dex. The run converged cleanly with zero divergences. The posterior summaries are
\begin{equation}
\logmphi=-19.20\pm0.10,
\qquad
\alpha=0.020\pm0.019.
\end{equation}
Figure~\ref{fig:robustness} compares the three free-\(\alpha\) runs. The three posterior summaries agree at the plotted precision. The widening applies only to the high-stellar-mass SMHM tier, \(\log_{10}\Mstar\geq9.5\). The boundary solution is stable under this prior-widening check.

\subsection{Fixed-\texorpdfstring{\(\alpha\)}{alpha} sensitivity sweep}

\begin{figure}[!htbp]
\centering
\includegraphics[width=0.76\linewidth]{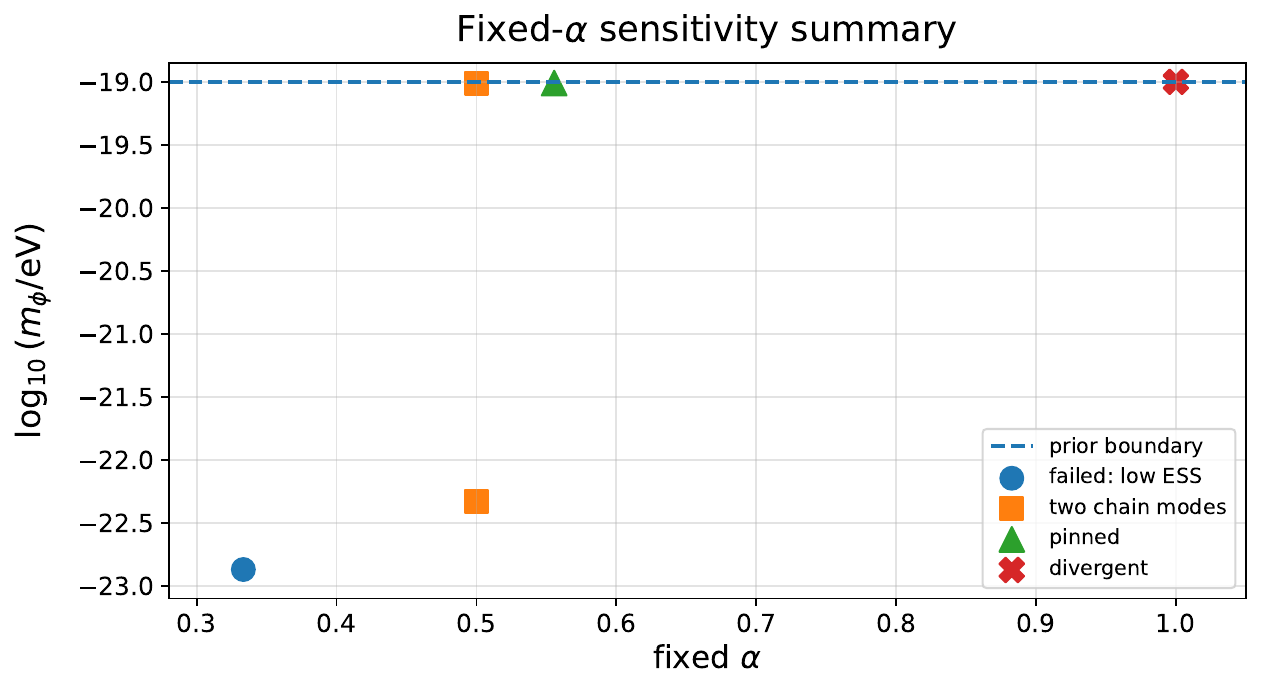}
\caption{Fixed-\(\alpha\) sensitivity summary. The \(\alpha=1/2\) run has two chain modes. The \(\alpha=5/9\) and \(\alpha=1.0\) runs sit at the high-mass prior boundary. The \(\alpha=1.0\) run has divergences. These runs probe posterior structure at fixed \(\alpha\) and are reported separately from the headline posterior.}
\label{fig:fixed_status}
\end{figure}

We ran four fixed-\(\alpha\) sensitivity runs at \(\alpha=1/3\), \(1/2\), \(5/9\), and \(1.0\). These runs probe posterior geometry away from the free-\(\alpha\) boundary solution. Table~\ref{tab:fixed_alpha} separates runs used as posterior results from runs reported only as sampler or geometry diagnostics. Figure~\ref{fig:fixed_status} gives a visual summary.

\begin{table}[!htbp]
\centering
\caption{Convergence diagnostics for the runs and the fixed-\(\alpha\) sensitivity sweep. ESS values are bulk and tail ESS for \(\logmphi\), except where a fixed-\(\alpha\) run is reported by chain behaviour because the chains occupy distinct modes.}
\label{tab:fixed_alpha}
\scriptsize
\setlength{\tabcolsep}{3pt}
\begin{tabularx}{\linewidth}{l c c c c c c Y}
\toprule
Run & \(\alpha\) & \(\logmphi\) & \(\rhat\) & ESS\(_{\rm bulk}\) & ESS\(_{\rm tail}\) & Div. & Verdict \\
\midrule
Free-\(\alpha\) baseline & free & \(-19.20\pm0.10\) & 1.000 & 1844 & 1582 & 0 & converged \\
Drop UGC06787 & free & \(-19.20\pm0.11\) & 1.001 & 2304 & 1680 & 0 & converged \\
SMHM \(\sigma=0.30\) & free & \(-19.20\pm0.10\) & 1.002 & 1687 & 1533 & 0 & converged \\
Fixed \(\alpha=1/3\) & 1/3 & \(-22.87\pm0.02\) & 1.30 & 29 & 12 & 0 & failed: \(\rhat=1.30\), low ESS \\
Fixed \(\alpha=1/2\) & 1/2 & two chain modes & -- & -- & -- & 0 & failed: chains in distinct modes near \(-19.01\) and \(-22.33\) \\
Fixed \(\alpha=5/9\) & 5/9 & \(-19.01\pm0.005\) & 1.000 & 5525 & 2195 & 0 & pinned at boundary \\
Fixed \(\alpha=1.0\) & 1.0 & \(-19.00\pm0.002\) & 1.07 & 35 & 164 & 22 & failed: divergences \\
\bottomrule
\end{tabularx}
\end{table}

The fixed-\(\alpha\) sweep does not provide a clean ridge measurement. It gives three concrete diagnostics: (i) the \(\alpha=1/3\) run finds an interior mass scale and fails the stated convergence criteria; (ii) the \(\alpha=1/2\) run is bimodal, with one chain near the high-mass boundary and one chain near an interior mode; (iii) the \(\alpha=5/9\) and \(\alpha=1.0\) runs move to the high-mass boundary, with divergences in the \(\alpha=1.0\) run. The fixed-\(\alpha\) values are imposed externally, so their \(\mphi\) summaries are conditional diagnostics rather than sample-level boson-mass measurements.

\FloatBarrier
\subsection{Per-galaxy halo-mass posteriors}

\begin{figure}[!ht]
\centering
\includegraphics[width=0.70\linewidth]{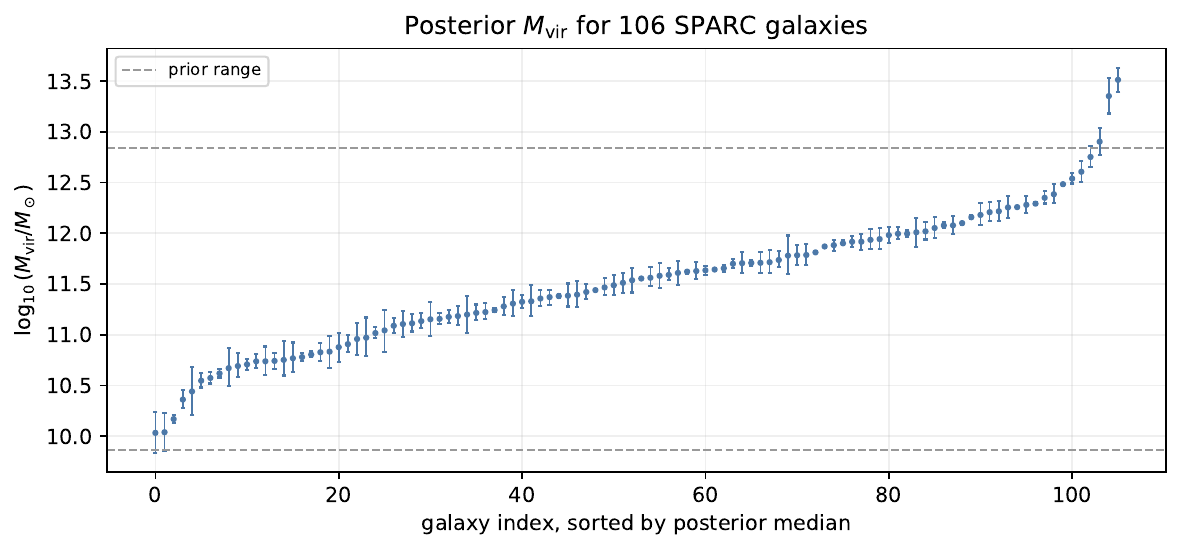}
\caption{Posterior \(\log_{10}(\Mvir/\Msun)\) across the 106-galaxy sample, sorted by posterior median. Error bars show the 16/84 percentile interval from the headline free-\(\alpha\) run. Dashed lines mark the overall halo-mass prior range induced by the sample. The local halo parameters are well behaved while the global soliton parameters move to the boundary solution.}
\label{fig:mvir}
\end{figure}

The per-galaxy \(\Mvir\) posteriors are well behaved under the headline run and are summarized in Fig.~\ref{fig:mvir}.

\FloatBarrier

The posteriors span the halo-mass range set by the \(V_{\rm flat}\) prior. The typical posterior width is about 0.2 dex. This shows that the local halo parameters are constrained by the combined velocity and halo-mass priors while the global soliton parameters move to the boundary solution.

\subsection{Representative rotation curve at the boundary solution}

\begin{figure}[H]
\centering
\includegraphics[width=0.78\linewidth]{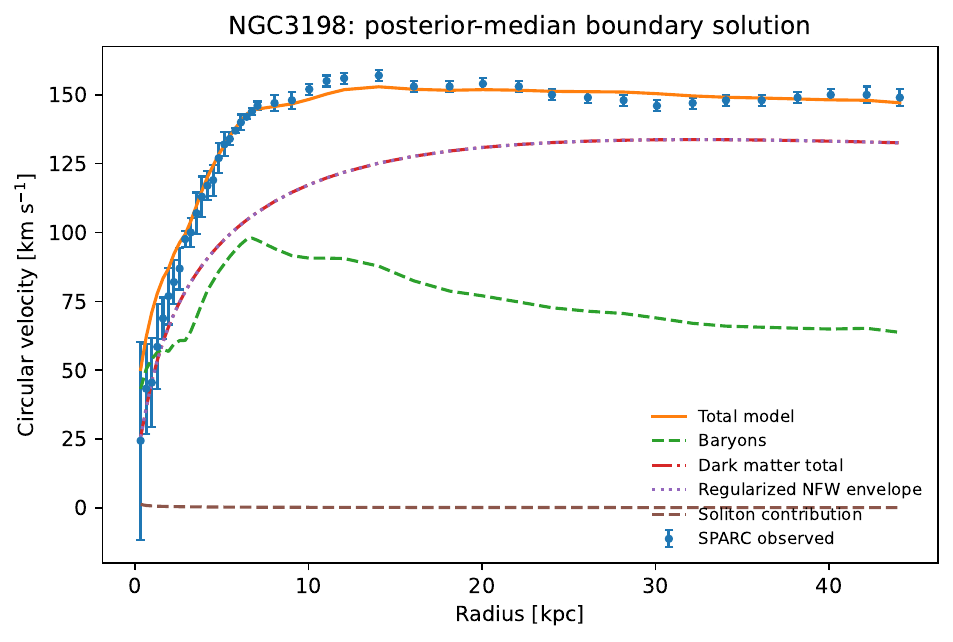}
\caption{Representative rotation-curve decomposition for NGC3198 using posterior-median parameters from the free-\(\alpha\) run. The observed SPARC points are compared with the total model, baryonic terms, total dark contribution, regularized NFW envelope, and soliton contribution. The soliton contribution is negligible across the observed radii; the visible rotation curve is carried by the baryonic terms and the regularized NFW envelope. This illustrates the compact-core interpretation of the boundary solution.}
\label{fig:rotation_ngc3198}
\end{figure}

Figure~\ref{fig:rotation_ngc3198} shows a representative rotation-curve decomposition for NGC3198 using posterior-median parameters from the free-\(\alpha\) run. The fit has \(\chi^2/N\simeq1.20\), which is a reasonable fit for a representative galaxy and is used here as an illustrative example rather than a sample-wide posterior-predictive check. The inferred core radius is \(\rc\simeq0.0023\kpc\), about 2.3 pc. This scale is far below the kiloparsec radii sampled by the SPARC rotation curve. The soliton contribution is negligible across the plotted radial range, and the baryonic terms plus the regularized NFW envelope carry the visible fit. This example illustrates the boundary-solution interpretation in velocity space.

\FloatBarrier

\section{Discussion}
\label{sec:discussion}

\subsection{What the model reports}

Under the density normalization in Eq.~\ref{eq:rhoc_schive} and the literature-standard prior \(\logmphi\in[-25,-19]\), the hierarchical SPARC fit returns a boundary solution. The free-\(\alpha\) posterior reaches the upper \(\mphi\) boundary and the lower \(\alpha\) boundary. The high-mass edge corresponds to the regime where the soliton becomes compact on observed rotation-curve scales. The low-\(\alpha\) edge corresponds to weak dependence of core radius on halo mass. In this regime, the regularized NFW envelope and baryonic terms carry the rotation-curve fit.

The limit \(\alpha\simeq0\) removes the inferred halo-mass dependence of the core radius. The hierarchy therefore does not recover a population-level core-halo scaling in the free-\(\alpha\) run. Combined with the high-\(\mphi\) boundary, this places the model in a compact-core regime across the observed SPARC radii. Inner-rotation-curve systematics can matter in this regime. Beam smearing, non-circular motions, and inclination uncertainties affect the same inner data points where a soliton contribution would be largest. A different treatment of those systematics could change the posterior.

The fixed-\(\alpha=1/3\) run gives an interior mass scale near \(\logmphi=-22.87\), although it fails the strict convergence criteria. This contrast is informative. An interior soliton solution can appear when the core-halo exponent is fixed. When \(\alpha\) is sampled, the hierarchy moves away from that conditional interior solution and reaches the boundary solution. The \(\alpha=1/2\) bimodality shows the same transition directly: one chain occupies an interior region and one chain occupies the high-mass boundary region. We therefore treat the fixed-\(\alpha\) runs as conditional probes of posterior structure.

The two robustness runs support the same conclusion. Removing UGC06787 leaves the posterior unchanged at the quoted precision. Widening the high-mass SMHM scatter leaves the posterior unchanged at the quoted precision. The boundary behaviour therefore appears across the tested free-\(\alpha\) configurations.

This result is specific to the modelling choices used here. It describes the joint behaviour of the selected SPARC sample, the forward model, and the hierarchical prior structure. A different selection, a different transition prescription, a different halo prior, or a different treatment of observational systematics may change the posterior. Several published nearby-galaxy FDM analyses report mass preferences under different assumptions \citep{Bar2018,Bar2022,Khelashvili2023,Banares2023}. A direct reconciliation requires matched data cuts, matched priors, and matched profile definitions.

\subsection{Relation to earlier SPARC FDM work}

\citet{Khelashvili2023} apply a Bayesian framework to SPARC galaxies with a hard soliton-NFW transition radius and per-galaxy mass preferences. They report tension between individual galaxies and conclude that no single \(\mphi\) fits the full sample. \citet{Banares2023} fit LITTLE THINGS dwarf irregulars with a fixed core-halo relation. \citet{Bar2022} use SPARC to test the soliton-halo relation as an exclusion criterion.

Our pipeline differs in three ways. It samples \(\alpha\) as a global parameter. It uses one global \(\mphi\) for the selected sample. It uses a differentiable composite profile with a smooth transition between the soliton and the regularized NFW envelope. Under these choices, the free-\(\alpha\) hierarchy moves to the high-mass, low-\(\alpha\) boundary region. This is consistent with the broad conclusion of \citet{Khelashvili2023} that a single FDM mass is difficult to identify across the SPARC sample, although the parameterizations and priors are not matched one-to-one.

The main difference is not only computational. By sampling \(\alpha\), the pipeline tests a population assumption that is otherwise fixed before inference. The contrast between the free- and fixed-\(\alpha\) runs shows why this matters. An imposed scaling can produce an interior mass mode, while the converged free-scaling hierarchy moves toward compact, unresolved cores. The framework therefore tests whether a particle-mass preference remains when the core-halo relation is allowed to respond to the data.

\subsection{What the priors are doing}

The \(V_{\rm flat}\) and Moster priors multiply as independent log-density terms and set the local halo-mass scale. The per-galaxy \(\Mvir\) posteriors concentrate inside the expected halo-mass range, as shown in Fig.~\ref{fig:mvir}. These priors keep the local halo parameters from drifting over several dex while the global soliton parameters move to the boundary solution.

The \(\mphi\) prior bound at \(-19\) has a different role. It marks the high-mass edge of the SPARC FDM prior range used by \citet{Khelashvili2023}. At this edge, the soliton radius becomes smaller than the observed rotation-curve scale for the selected galaxies. The free-\(\alpha\) posterior reaches this edge. The result therefore reports the limit of this model on the selected SPARC sample: the inference moves toward the compact-core, weak-halo-dependence regime instead of an interior soliton solution.

\subsection{Use of the pipeline beyond the present sample}

The pipeline is not tied to the 106 galaxies used here. The data layer, population parameters, galaxy-level parameters, and density model are separate parts of the implementation. A new rotation-curve sample can be supplied without changing the basic hierarchy. The soliton and envelope prescriptions can also be replaced while keeping the same sampling and diagnostic structure.

Synthetic populations are a direct use case. Mock rotation curves can be generated with a known particle mass and core-halo exponent and passed through the same selection, likelihood, and inference steps. Such injection-recovery tests can show when the hierarchy recovers an interior soliton signal and when it moves toward the unresolved-core limit. They can also separate the effects of sample size, inner radial coverage, velocity uncertainty, and the assumed halo relation.

Larger samples provide more population information, but sample size alone is not enough. The present boundary solution places the soliton below the radii probed by the selected rotation curves. Future applications will be most informative when additional galaxies are combined with well-resolved inner kinematics and controlled distance, inclination, and non-circular-motion uncertainties. Under those conditions, the joint hierarchy can test whether the particle mass and scaling relation become identifiable together.

\subsection{Limitations and future work}

The present application is a first test of the joint-inference pipeline. Its scope is the posterior behaviour of one soliton-plus-NFW hierarchy, not a ranking of all halo models. Within that scope, the main result is stable: the free-\(\alpha\) posterior moves to the low-\(\alpha\), high-\(\mphi\) boundary solution under the completed checks.

First, the analysis does not include simulation-based recovery tests on synthetic SPARC-like data. Such tests would be useful for checking how the hierarchy responds to an injected interior soliton component. Second, the paper does not present a full model-comparison study against soliton-free halo models. The focus here is the posterior behaviour of the soliton-plus-NFW hierarchy rather than a ranking of halo prescriptions. Third, the representative NGC3198 decomposition in Fig.~\ref{fig:rotation_ngc3198} is not a replacement for a sample-wide posterior-predictive analysis.

Several modelling approximations should also be kept in mind. The \(V_{\rm flat}\) prior uses a fixed reference concentration, \(c_{200}=10\), while the model samples \(c_{200}\) separately. The SMHM scatter tier is assigned using the prior-mean disk mass-to-light ratio \(\Upsilon_{\rm disk}=0.5\), rather than the sampled value. For galaxies with posterior \(\Upsilon_{\rm disk}\) far from the prior mean and stellar mass near a tier boundary, the assigned tier may differ from the tier that would be chosen using the sampled value.

The fixed-\(\alpha\) runs at \(1/3\), \(1/2\), and \(1.0\) require improved sampling before quantitative use. The widened-prior diagnostic in Appendix~\ref{app:widened} also did not converge under the baseline sampler settings. These runs are therefore treated as diagnostics of posterior geometry, not as posterior measurements.

The next step is to use the framework for synthetic-data recovery tests. These tests can determine which soliton signals are recoverable for SPARC-like radial sampling and uncertainties. Soliton-free baseline comparisons and sample-wide posterior-predictive checks can then test whether a resolved soliton improves the fit when one is present. Further applications can include distance and inclination marginalization, longer warmup tests, alternative transition prescriptions, and larger or higher-resolution rotation-curve samples.

\section{Conclusions}
\label{sec:conclusion}

We presented a reusable hierarchical Bayesian framework that infers the ultralight particle mass and the population-level core-halo exponent jointly from galaxy rotation curves. The main change from a fixed-scaling analysis is that the galaxy sample is allowed to determine both global quantities. The pipeline separates these global dark-matter parameters from the halo and stellar parameters of each galaxy and uses a differentiable composite profile so that the full model can be sampled with Hamiltonian Monte Carlo.

The findings were:.
\begin{enumerate}
\item The free-\(\alpha\) run on 106 SPARC galaxies converges cleanly with zero divergences, \(\rhat\simeq1.000\), and bulk effective sample sizes above 1800 for the global parameters.
\item The posterior medians and 16/84 percentile intervals are \(\logmphi=-19.20^{+0.12}_{-0.11}\) and \(\alpha=0.014^{+0.023}_{-0.011}\). The posterior reaches the high-\(\mphi\) and low-\(\alpha\) prior boundaries, so boundary fractions and quantiles carry the primary interpretation.
\item Removing UGC06787 and widening the high-mass SMHM prior both return the same boundary solution.
\item Fixed-\(\alpha\) sensitivity runs reveal difficult posterior geometry away from the free-\(\alpha\) solution. Most fixed-\(\alpha\) runs either fail convergence checks or sit at the high-mass prior boundary.
\end{enumerate}

The present SPARC application shows what the joint framework reports when the proposed population structure is not resolved by the data. Within the adopted Schive-normalized model and standard FDM prior range, the hierarchy does not identify an interior particle mass and core-halo scaling relation. It moves instead to a compact-core limit in which the soliton lies below the observed radial scales. This prevents the boundary value from being misread as an interior particle-mass measurement.

The same framework can now be used to ask when a joint constraint becomes possible. Synthetic populations can test recovery of known particle masses and scaling relations. Expanded or higher-resolution rotation-curve samples can test whether better population statistics and inner radial coverage resolve the soliton component. Alternative core and envelope models can be inserted into the same hierarchy. The code, run manifests, posterior samples, and diagnostic tables are provided so that these tests can be carried out without rebuilding the inference pipeline.

\FloatBarrier
\begin{appendices}

\section{Full priors and hyperparameters}
\label{app:priors}

This appendix records the prior configuration used for the reported runs. Table~\ref{tab:priors} gives the global priors, galaxy-level priors, and hyperparameter choices used by the sampler.

\begin{table}[!htbp]
\centering
\caption{Full prior specification.}
\label{tab:priors}
\tiny
\renewcommand{\arraystretch}{0.88}
\setlength{\tabcolsep}{2pt}
\begin{tabularx}{\linewidth}{p{0.23\linewidth} p{0.34\linewidth} Y}
\toprule
Parameter & Prior or value & Source or role \\
\midrule
\(\logmphi\) & \(U(-25,-19)\) & SPARC FDM prior range following \citet{Khelashvili2023} \\
\(\alpha\) & \(U(0,1.5)\) & broad positive core-halo exponent range \\
\(A_{\rm ch}\) & \(1.6\kpc\) & core-radius normalization from \citet{Schive2014PRL} \\
\(\rho_{\rm norm}\) & \(1.9\times10^7\,\Msun\,\mathrm{kpc}^{-3}\) & Schive soliton density normalization \\
\(k\) & 3 & logistic blending steepness \\
Transition search & soft-argmin with \(\beta=12\) & density-intersection search on a log-radius grid \\
Mass integration grid & 256 log-spaced + 320 linear points & numerical enclosed-mass integration \\
\(\log_{10}\Mvir^{(i)}\) & \(\mathcal{N}(\mu^{(i)}_{V_{\rm flat}},0.35)\) & \(V_{\rm flat}\)-based halo-mass prior \\
\(\log_{10}c_{200}^{(i)}\) & \(\mathcal{N}(0.905-0.101[\log_{10}(\Mvir/\Msun)-12],0.11)\) & mass-concentration relation from \citet{Dutton2014} \\
\(\Upsdisk^{(i)}\) & log-normal centered on 0.5, width 0.20 dex & stellar mass-to-light prior \\
\(\Upsbul^{(i)}\) & log-normal centered on 0.7, width 0.15 dex & bulge mass-to-light prior \\
SMHM scatter, high \(\Mstar\) & 0.20 dex & \(\log_{10}\Mstar\geq9.5\) \\
SMHM scatter, mid \(\Mstar\) & 0.30 dex & \(9.0\leq\log_{10}\Mstar<9.5\) \\
SMHM prior, low \(\Mstar\) & disabled & avoids dwarf-regime extrapolation \\
Luminosity floor & \(10^5L_\odot\) & makes SMHM prior effectively uninformative when photometry is unavailable \\
\bottomrule
\end{tabularx}
\end{table}

These choices define the hierarchy used throughout the analysis. The sensitivity runs in the main text change only the parameter stated for that run.

\FloatBarrier
\section{Per-galaxy diagnostics and fixed-\texorpdfstring{\(\alpha\)}{alpha} traces}
\label{app:fixed_traces}

\begin{figure}[!htbp]
\centering
\includegraphics[width=0.95\linewidth]{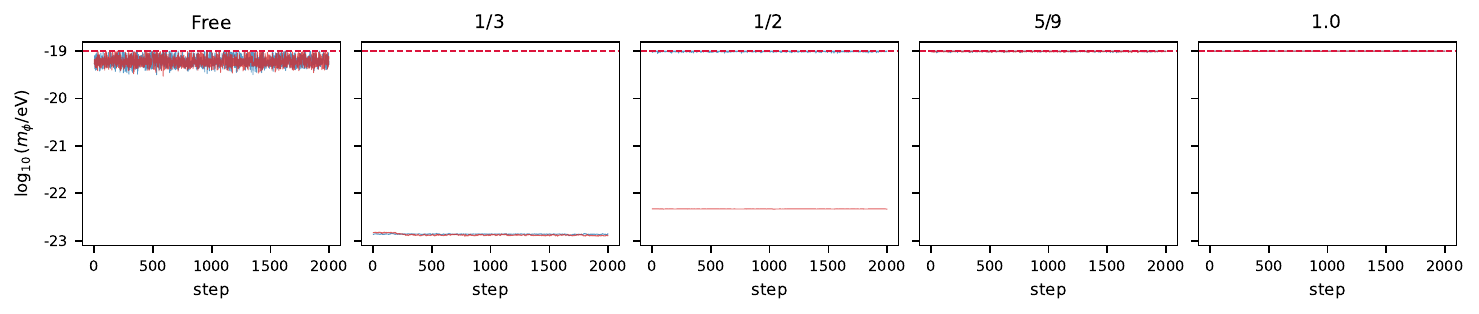}
\caption{Trace plots for the free-\(\alpha\) baseline and fixed-\(\alpha\) sensitivity runs. The red dashed line marks \(\logmphi=-19\).}
\label{fig:fixed_traces}
\end{figure}

The per-galaxy halo-mass posteriors are summarized in Fig.~\ref{fig:mvir}. The fixed-\(\alpha\) trace plots are reproduced in Fig.~\ref{fig:fixed_traces}. The \(\alpha=1/3\) chains drift slowly. The \(\alpha=1/2\) chains occupy distinct modes. The \(\alpha=5/9\) and \(\alpha=1.0\) chains sit near the high-mass prior boundary. The \(\alpha=1.0\) run has divergences. These traces support the status labels in Table~\ref{tab:fixed_alpha}.

\FloatBarrier
\setcounter{table}{0}
\renewcommand{\thetable}{C\arabic{table}}
\section{Widened-prior diagnostic}
\label{app:widened}

We ran one diagnostic configuration with the upper \(\mphi\) prior bound extended from \(-19\) to \(-15\). This configuration produced 2754 divergences out of 4000 post-warmup samples, \(\rhat=1.41\) for \(\logmphi\), \(\rhat=1.89\) for \(\alpha\), and \(\mathrm{ESS}_{\rm bulk}(\alpha)=2.9\). The chains explored different regions of the high-\(\mphi\) space and did not agree. The configuration did not produce a reliable posterior under the baseline sampler settings.

This run is reported as a sampler diagnostic. It indicates difficult sampling geometry in the high-mass region of the forward model. At masses well above the standard upper bound, the inferred core radius can become far smaller than the observed rotation-curve scale, which makes the forward model physically ill-conditioned for rotation-curve inference. The run does not enter any posterior summary.

\begin{table}[!htbp]
\centering
\caption{Widened-prior diagnostic with \(\logmphi\in[-25,-15]\). The run did not converge and is shown only as a sampler diagnostic.}
\label{tab:widened}
\begin{tabular}{ll}
\toprule
Quantity & Value \\
\midrule
Configuration & \(\logmphi\in[-25,-15]\) \\
Divergences & 2754 / 4000 post-warmup samples \\
\(\rhat(\logmphi)\) & 1.41 \\
\(\rhat(\alpha)\) & 1.89 \\
\(\mathrm{ESS}_{\rm bulk}(\alpha)\) & 2.9 \\
Status & non-converged diagnostic \\
\bottomrule
\end{tabular}

\vspace{2pt}
{\small This configuration is not used as a posterior result.}
\end{table}

\FloatBarrier
\section{Code and data availability}
\label{app:code}

The pipeline is implemented in Python with JAX and NumPyro. The codebase, run configuration files, SPARC selection manifest, posterior sample archives, and diagnostic tables are available at \url{https://github.com/arham766/uldm-sparc-hbi}. The SPARC data are public \citep{Lelli2016}. 

\end{appendices}

\bmhead{Acknowledgements}
The authors thank Dr. Dennis Krause for mentoring the project and reviewing the analysis. The authors thank Dr. Nathan Tompkins for his support and for accepting this work as an Advanced Lab project. The authors thank the Wabash College Department of Mathematics and Computer Science for access to computing resources.

\bmhead{Author contributions}
Prasun Panthi led the theoretical framing, model interpretation, and manuscript preparation. Md Shahrier Islam Arham led the computational implementation, run orchestration, and diagnostic packaging.

\bmhead{Funding}
No external funding was received for this work.

\bmhead{Competing interests}
The authors declare no competing interests.

\bibliography{references}

\end{document}